\documentclass[aps,prd,onecolumn,nofootinbib,amsmath,amssymb]{revtex4}

\usepackage{subfig}
\usepackage{hyperref}
\usepackage{epsfig}
\usepackage{graphicx}

\begin{document}

\title{A search for inverse magnetic catalysis in thermal quark-meson models}

\author{E.~S.~Fraga$^{1,2,3}$, B.~W.~Mintz$^{4}$ and J.~Schaffner-Bielich$^{1}$}

\affiliation{$^{1}$Institute for Theoretical Physics, Goethe University, D-60438 Frankfurt am Main, Germany\\
$^{2}$Frankfurt Institute for Advanced Studies, Goethe University, D-60438 Frankfurt am Main, Germany\\
$^{3}$Instituto de F\'\i sica, Universidade Federal do Rio de Janeiro,
Caixa Postal 68528, 21941-972, Rio de Janeiro, RJ , Brazil\\
$^{4}$Departamento de F\'{\i}sica Te\'orica, Universidade do Estado 
do Rio de Janeiro, 20550-013, Rio de Janeiro, RJ, Brazil}
%


\begin{abstract}
  We explore the parameter space of the two-flavor thermal quark-meson
  model and its Polyakov loop-extended version under the influence of
  a constant external magnetic field $B$. We investigate the behavior
  of the pseudo critical temperature for chiral symmetry breaking
  taking into account the likely dependence of two parameters on the
  magnetic field: the Yukawa quark-meson coupling and the parameter
  $T_0$ of the Polyakov loop potential. Under the constraints that
  magnetic catalysis is realized at zero temperature and the chiral
  transition at $B=0$ is a crossover, we find that the quark-meson
  model leads to thermal magnetic catalysis for the whole allowed
  parameter space, in contrast to the present picture stemming from
  lattice QCD.
%
\end{abstract}

\maketitle


\section{Introduction}

The phase diagram of magnetic quantum chromodynamics QCD, i.e.\ for
strong interactions under the influence of an external classical
Abelian magnetic field, is currently under construction. This phase
diagram corresponds to a special case, as it does not suffer from the
Sign Problem, and can be easily simulated on the lattice. So, from
the theoretical point of view, it serves as a crucial check for
effective models of QCD extended to regions not easily accessible by lattice
simulations. From the experimental standpoint, this setup is also
quite remarkable owing to the fact that strong magnetic fields are
relevant in non-central heavy ion collisions, and play a major role in
the possibility of observing the chiral magnetic effect (for a
comprehensive review, see Ref.~\cite{Kharzeev:2013jha}.)

The chiral and deconfining transitions under the effect of a magnetic
background are, of course, amenable also to effective model
descriptions \cite{Fraga:2012rr,Gatto:2012sp}. Those models have
predicted several outstanding new features to the thermodynamics of
strong interactions, from shifting the chiral and the deconfinement
crossover lines in the phase diagram
\cite{Agasian:2008tb,Fraga:2008qn,Menezes:2008qt,Boomsma:2009yk,Mizher:2010zb,Fukushima:2010fe,Gatto:2010pt,Kashiwa:2011js,Chatterjee:2011ry,Andersen:2011ip,Skokov:2011ib,Andersen:2012bq,Fukushima:2012xw,Fraga:2012fs,Strickland:2012vu,Fukushima:2012kc,Ferrari:2012yw,Allen:2013eha,Fu:2013ica,Andersen:2012jf,Andersen:2012zc,Avancini:2012ee,Andersen:2012dz,Costa:2013zca,Allen:2013lda,Ballon-Bayona:2013cta,Callebaut:2013ria,Ferreira:2013tba,Ruggieri:2013cya}
to transforming the vacuum into a superconducting medium via
$\rho$-meson condensation \cite{Chernodub:2010qx,Chernodub:2011mc},
for high enough magnetic fields, i.e.\ a few times
$m_{\pi}^2$. Nevertheless, the available lattice data
\cite{D'Elia:2010nq,D'Elia:2011zu,Bali:2011qj,Bali:2012zg,Bruckmann:2013oba}
contradicted essentially all predictions regarding the behavior of the
pseudo critical lines for deconfinement and chiral symmetry
restoration coming from chiral models (including their Polyakov loop
extensions).  The reason for this failure is unclear, but the fact
that confinement is not properly captured in such chiral models might
play a role \cite{Fraga:2012fs,Fraga:2012ev}, as well as an apparently
nontrivial chiral limit \cite{Blaizot:2012sd}.

In this paper, we explore the parameter space of the two-flavor
thermal quark-meson (QM) model and its Polyakov loop-extended version
(PQM) under the influence of a constant external magnetic field
$B$. We investigate the behavior of the pseudo critical temperature
for chiral symmetry breaking taking into account the likely dependence
of two parameters on the magnetic field: the Yukawa quark-meson
coupling and the parameter $T_0$ of the Polyakov loop potential. We
scan an important part of the parameter spaces of these models, in
order to check whether they can accommodate, at least qualitatively,
the trend of inverse magnetic catalysis found in
Ref. \cite{Bali:2011qj}. In doing so, we keep two important
constraints: (i) that magnetic catalysis is realized at zero
temperature, and (ii) that the chiral transition at $B=0$ is a
crossover. In these two limits, i.e.\ zero temperature or zero magnetic
field, chiral effective models usually produce results that are, at
least qualitatively, in line with lattice QCD. We find, nevertheless,
that the extensions considered by introducing a $B$ dependence in the
Yukawa coupling and in the parameter $T_0$ are not enough to account
for the behavior of the critical temperature as found in lattice QCD
simulations.  We believe this should be also the case for other chiral
models, which would signal the lack of some fundamental ingredient,
possibly quark confinement.

The paper is organized as follows. In Section II we describe the
thermal effective potential in the presence of a magnetic background.
In Section III we consider the effects from the running of the Yukawa
coupling, having the magnetic field as the momentum scale, in the QM
model. In Section IV we investigate the consequences of making $T_0$
$B$ dependent, in a fashion usually made with chemical
potentials. Section V contains our summary.

\section{Thermal effective potential in the presence of a magnetic field}
%
Let us first briefly review the Polyakov-Quark-Meson (PQM) model while
we introduce our notation.  The effective potential of the PQM model
at finite temperature and magnetic field was calculated in
\cite{Mizher:2010zb}. At the mean-field level, it is a sum of four
contributions,
\begin{eqnarray}\label{eq:Veff-normalized}
 V_{eff}(\sigma,\phi_1,\phi_2,B,T) = V_{cl}(\sigma) + V_P(\phi_1,\phi_2,T,T_0) 
+ V_{vac}(\sigma,B) + V_{para}(\sigma,\phi_1,\phi_2,B,T),
\end{eqnarray}
where 
\begin{equation}\label{eq:mesonic-potential}
 V_{eff}(\sigma) = \frac{\lambda}{4}(\sigma^2-v^2) - h\sigma
\end{equation}
is the tree-level potential for the $\sigma$ field. The Polyakov loop
potential (in the logarithmic parametrization) is given by
\cite{Roessner_08}
\begin{equation}\label{eq:VPolyakov}
 \frac{V_P(\phi_1,\phi_2,T,T_0)}{T^4} = 
-\frac{a(T)}{2}L^*L+b(T)\log\left[1-6L^*L+4(L^{*3}+L^3)-3(L^*L)^2\right].
\end{equation}
The $T=0$ contribution to the effective potential (with the
normalization $V_{vac}(\sigma,0)\equiv0$) has the form
\begin{equation}\label{eq:VvacB}
 V_{vac}(\sigma,B) =
 -\frac{N_c}{2\pi^2}\sum_{f=u,d}|q_fB|^2\left[\zeta'(-1,x_f) 
-\frac{1}{2}(x_f^2-x_f)\log x_f+\frac{x_f^2}{4}\right],
\end{equation}
whereas $V_{para}(\sigma,\phi_1,\phi_2,B,T)$ is given by
\begin{eqnarray}\label{eq:VparaB}
 V_{para} = -\frac{BT}{\pi^2}\sum_{f=u,d}\sum_{n=0}^\infty(2-\delta_{n,0})\sum_{i=1}^3 
 \int_0^\infty dp
    \log\left(1+2e^{-\sqrt{p^2+g^2\sigma^2+2n|q_fB|}/T}\cos\phi_i + e^{-2\sqrt{p^2+g^2\sigma^2+2n|q_fB|}/T}\right),
\end{eqnarray}
and represents the paramagnetic (thermal) contribution to the effective potential. 

In the previous formulas, $B$ is the magnetic field, $T$ is the
temperature, $\sigma$ is the expectation value of the sigma meson
field (the approximate order parameter for the chiral transition),
$\phi_1$, $\phi_2$, and $\phi_3$ are the (approximate) deconfinement order
parameters related to the Polyakov loop $L$ (see, e.g.,
\cite{Mizher:2010zb} for more details).  The quantities $\lambda$, $v$
and $h$ are parameters of the mesonic self interaction
(\ref{eq:mesonic-potential}), which are adjusted according to the
constants on Table \ref{tab1:chiral_pot_constants}, and the functions
$a(T)$ and $b(T)$ are defined as
\begin{equation}
 a(T) = a_0 + a_1\left(\frac{T_0}{T}\right) + a_2\left(\frac{T_0}{T}\right)^2\;\;\;\; ; \;\;\;\; 
               b(T)=b_3\left(\frac{T_0}{T}\right)^3,
\end{equation}
where the values of the parameters $a_0$, $a_1$, $a_2$, and $b_3$ are
listed in Table \ref{tab2:ploop_pot_constants}.  In eq.~(\ref{eq:VvacB}),
we used $x_f:=m_f^2/|q_fB|=g^2\sigma^2/|q_fB|$, where $g$ is the
meson-quark coupling and $q_f$ is the charge of the quark of flavor
$f=u,d$.
\begin{table}
	\caption{Values of constants to which the parameters of the
          mesonic potential are adjusted, according to
          Ref.~\cite{PDG_12}.
 }
		\begin{tabular}{l||c|c|c|c|c|c|c}
		\hline\noalign{\smallskip}
                         Constant & $f_\pi$ [MeV] & $m_\pi$ [MeV] & $\lambda$ & $g_0$ & $m_f$ [MeV] & $q_u$   & $q_d$
                \\ \hline  Value  & 93            & 138           & 20        & 3.3   & 307         & $+1/3$  & $-2/3$ \\
                \noalign{\smallskip}\hline
		\end{tabular}
	\label{tab1:chiral_pot_constants}
\end{table}

\begin{table}
	\caption{Values of the parameters of the Polyakov loop potential \cite{Roessner_08}.
 }
		\begin{tabular}{l||c|c|c|c|c|c|c|c}
		   \noalign{\smallskip}\hline
                         Constant & $a_0$                   & $a_1$ & $a_2$ & $b_3$ & $T_0$ [MeV]
                \\ \hline  Value  & $16\pi^2/45\simeq3.51$  & $-2.47$ & $15.2$  & $-1.75$ & $270$ \\
       \noalign{\smallskip}\hline
		\end{tabular}
	\label{tab2:ploop_pot_constants}
\end{table}

Notice that the Quark-Meson (QM) model is recovered, as a particular
case of the PQM model, when one sets $\phi_1=\phi_2=\phi_3=0$ (or,
equivalently, $L=L^*=1$) and neglects the Polyakov loop potential
($V_P\equiv0$). In the PQM model, it can be shown that the reality of
the minimum of the effective potential implies
$\phi_3=-(\phi_1+\phi_2)=0$, so that there are actually only two order
parameters in the model, $\sigma$ and $\phi\equiv\phi_1$.

The state of thermodynamic equilibrium is the minimum of the effective
potential with respect to the order parameters $\sigma$ and $\phi$ at
fixed external conditions ($T$ and $B$). One defines, then, the chiral
condensate $\sigma(B,T)$, as well as the chiral susceptibility
$\chi(B,T)=\partial\sigma(B,T)/\partial T$.  The pseudo critical
temperature for chiral symmetry breaking is identified as the peak of
the chiral susceptibility $\chi(B,T)$.  As discussed in
Ref. \cite{Mizher:2010zb}, the pseudo critical temperature for the
chiral and the deconfinement transitions split as the magnetic field
increases sufficiently, a fact that is not in line with lattice
results \cite{Bali:2011qj}. To keep our discussion simpler, we will
focus our analysis on the chiral transition.

In the next two sections we promote two parameters to functions of the
external magnetic field: the Yukawa coupling $g$ and the parameter
$T_0$ of the Polyakov loop potential.


\section{B-dependent Yukawa coupling in the QM model}

\begin{figure}[!hbt]
\begin{center}
\includegraphics[width=9cm]{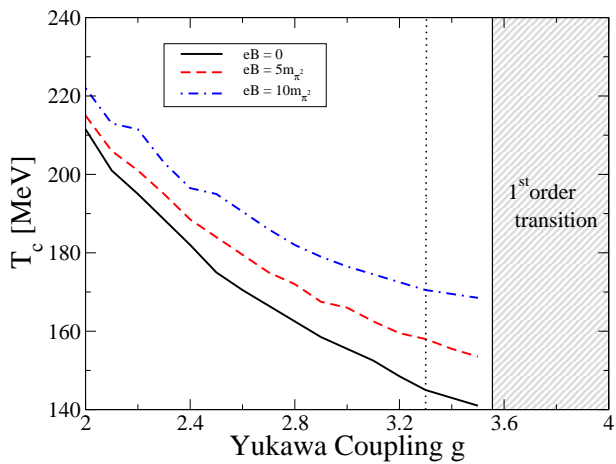}
\caption{$T_c$ as a function of the coupling $g$ for various values of
  $B$. Black (continuous) line: $eB=0$. Red (dashed) line:
  $eB=5m_\pi^2$. Blue (dash-dotted) line: $eB=10m_\pi^2$. Couplings
  higher than $g_c\simeq3.6$ lead to a first-order phase transition
  and are not shown in the plot. The vertical dotted line represents
  the value $g_0=3.3$, the physical value for $g$ in the vacuum. The
  vertical solid line corresponds to $g=3.6$.  }
\label{fig1:Tc-vs-g_severalB}
\end{center}

\end{figure}

Let us first consider the modifications brought about by introducing a
$B$ dependence in the Yukawa coupling, i.e.\ by considering
$g=g(B)$. For that purpose, we adopt the QM model (which is equivalent
to setting $\phi_1=\phi_2=\phi_3=0$ and $V_P\equiv0$ in the PQM
model).  This will change our formulas (\ref{eq:VvacB}) and
(\ref{eq:VparaB}) for the effective potential simply by the
replacement $g\rightarrow g(B)$. In order to describe the parameter
space of $g(B)$, we show the pseudo critical temperature as a function
of the Yukawa coupling $g$ for different values of the applied field
$B$, as seen in Fig. \ref{fig1:Tc-vs-g_severalB}.

Fig. \ref{fig1:Tc-vs-g_severalB} should be understood in the following
way. Each continuous function $g(B)$ will correspond to some
continuous path on the plot. All these paths must start on the
continuous line ($eB=0$), then proceed to some point on the dashed
line ($eB=5m_\pi^2$), then to the dotted line ($eB=10m_\pi^2$), and so
on, as $B$ increases. It is clear from
Fig. \ref{fig1:Tc-vs-g_severalB} that any choice $g(B)=g_0$ (a
constant) leads to an increasing $T_c(B)$, i.e.\ magnetic
catalysis. The same behavior will set in if $g(B)$ is a decreasing
function of $B$. The only possible way to find a decreasing $T_c(B)$
would be to let $g(B)$ {\it increase} with $B$.
 
However, as can be seen from Fig.  \ref{fig1:Tc-vs-g_severalB}, such a
function would have to grow very strongly. Thus, the coupling $g$
would very soon reach the region $g\gtrsim3.6$, where the transition
becomes of first order. Lattice results \cite{Bali:2011qj} clearly
show that the transition is a crossover for finite $B$ and, therefore,
we do not consider first-order transitions as acceptable solutions in
this paper. Besides, if $g(0)=3.3$, as required by the parameter fit
in the vacuum ($B=0$ and $T=0$), there is clearly no continuous $g(B)$
that would give rise to a decreasing $T_c(B)$ (at least not for a
crossover, see Fig. \ref{fig1:Tc-vs-g_severalB}). Therefore, we can
conclude that, {\it if one takes the usual parameter fixing in the
  vacuum, $g(0)=3.3$, there is no continuous function $g(B)$ that
  could lead to inverse magnetic catalysis in the QM model at finite
  temperature and zero quark chemical potential, unless the chiral
  transition is of first order.}

\begin{figure}[!hbt]
\begin{center}
\includegraphics[width=9cm]{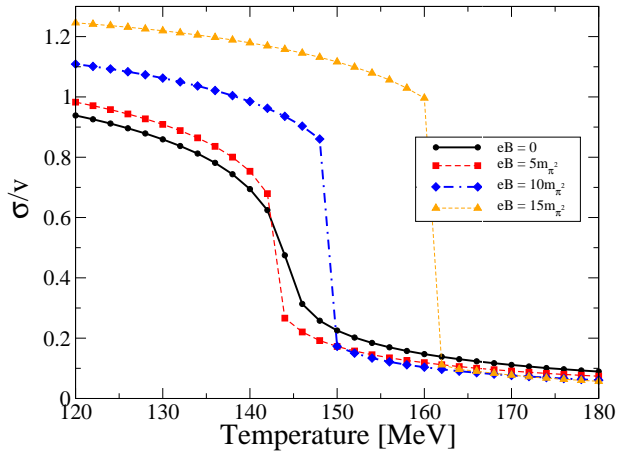}
\caption{Evolution of the normalized order parameter $\sigma/v$ as a
  function of the temperature for a B-dependent Yukawa coupling
  $g(B)=g_0(1+0.01eB/v^2)$. Notice the first-order phase transition at
  high $B$ (corresponding to $g>3.6$).}
\label{fig2:sigma_vs_T}
\end{center}
\end{figure}

To illustrate the general trend discussed above, we consider two
explicit examples. We show in Fig.~\ref{fig2:sigma_vs_T} the
evolution of the chiral order parameter $\sigma(B,T)$ as a function of
the temperature for different values of $B$. We take the (increasing)
ansatz $g(B)=g_0(1+0.01eB/v^2)$. For small values of the applied
field, the transition is a crossover. However, as soon as $B$ reaches
a moderate value ($eB\lesssim5m_\pi^2$), the chiral crossover turns
into a first-order transition.

\section{B-dependent $T_0$ in the PQM model}

Let us now consider the PQM model with the standard (fixed) Yukawa
coupling $g=g_0=3.3$, but with a varying $T_0=T_0(B)$.  As discussed
in Ref. \cite{Schaefer_07,Roessner_08}, the model parameter $T_0$ is 
equal to the deconfinement critical temperature in a pure gauge Polyakov 
loop model, $T_0^{gauge}=270$ MeV. However, due to the back reaction 
from the matter fields, $T_0$ should be lower than its pure gauge 
value when dynamical quarks are present. More
specifically, one takes $T_0$ as the running scale in the perturbative
renormalization group running of $\alpha_s$, the strong coupling
constant. Once $\alpha_s$ depends also on the number of quark flavors,
quark masses, and possibly on external constraints (e.g. on the
chemical potential, as in Ref. \cite{Schaefer_07}, or on the magnetic
field, as considered here), a function
$T_0(N_f,m_f,B)=T_\tau\exp\left[-1/\alpha_sb(N_f,m_f,B)\right]$ may be
implemented, where $T_\tau$ is a given energy scale used to fix
$\alpha_{s}$ (e.g., the tau lepton mass $m_\tau=1777$GeV \cite{PDG_12}).

\begin{figure}[!hbt]
\includegraphics[width=9cm]{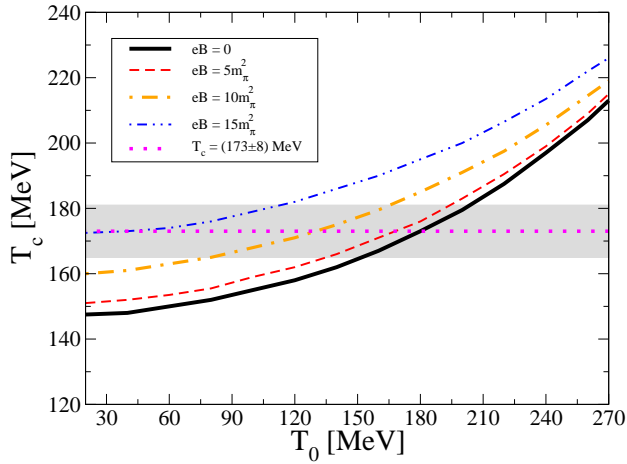}
 \caption{Pseudo critical temperature $T_c$ as a function of $T_0$ for 
 several values of the applied magnetic field in the PQM model. The 
 band $T_c=(173\pm8)$MeV corresponds to the pseudo critical temperature 
 found on the lattice for $N_f=2$ \cite{Karsch:2000kv}.
 }
\label{fig3:t0-vs-B}
\end{figure}

According to the analysis of Ref. \cite{Bruckmann:2013oba,Bruckmann:2013ufa}, 
the form of the Polyakov loop 
effective potential could be important for the mechanism of 
inverse magnetic catalysis around the transition temperature. 
A natural way to parametrize the $B$ dependence of the Polyakov 
loop potential is through the $T_0$ parameter, described above. 
We investigate here whether or not this possibility can lead to
inverse magnetic catalysis. We find that,
within the PQM model (with all other parameters kept at their vacuum
values), no function $T_0(B)$ can lead to a decreasing 
$T_c(B)$ for sufficiently high magnetic fields, such as those investigated 
in \cite{Bali:2011qj}. Some choices for $T_0(B)$ may, at 
most, make the pseudo critical temperature decrease for {\it small} 
values of
$B$. {\it However, even for magnetic field intensities $eB\sim0.4~
  GeV^ 2\sim 20 m_\pi^ 2$, the catalysis-inducing vacuum term
  (\ref{eq:VvacB}) dominates the whole picture and $T_c(B)$ inevitably
  rises.}

In order to see that the pseudo critical temperature tends to rise for
every parametrization $T_0(B)$, we proceed in the same fashion as in
the analysis of $g(B)$ in the previous section. Fig. \ref{fig3:t0-vs-B} 
shows, for different intensities of magnetic field, how the pseudo critical 
temperature $T_c$ depends on the value of $T_0$. For reference, we also show 
the value $T_c(B=0)=(173\pm8)$MeV, obtained in the two-flavor lattice calculation 
of \cite{Karsch:2000kv}. 
In this diagram, each continuous function $T_0(B)$ 
corresponds to some line connecting the curves shown. One can immediately 
realize that the pseudo critical temperature $T_c(B)$ may go down only 
if $T_0(B)$ decreases sufficiently fast. Such behavior, however, can not 
be sustained even at moderate intensities of the magnetic field, regardless 
of how fast $T_0(B)$ may decrease as a function of $B$. 

\begin{figure}
 \centering
 \subfloat[]{\includegraphics[width=7cm]{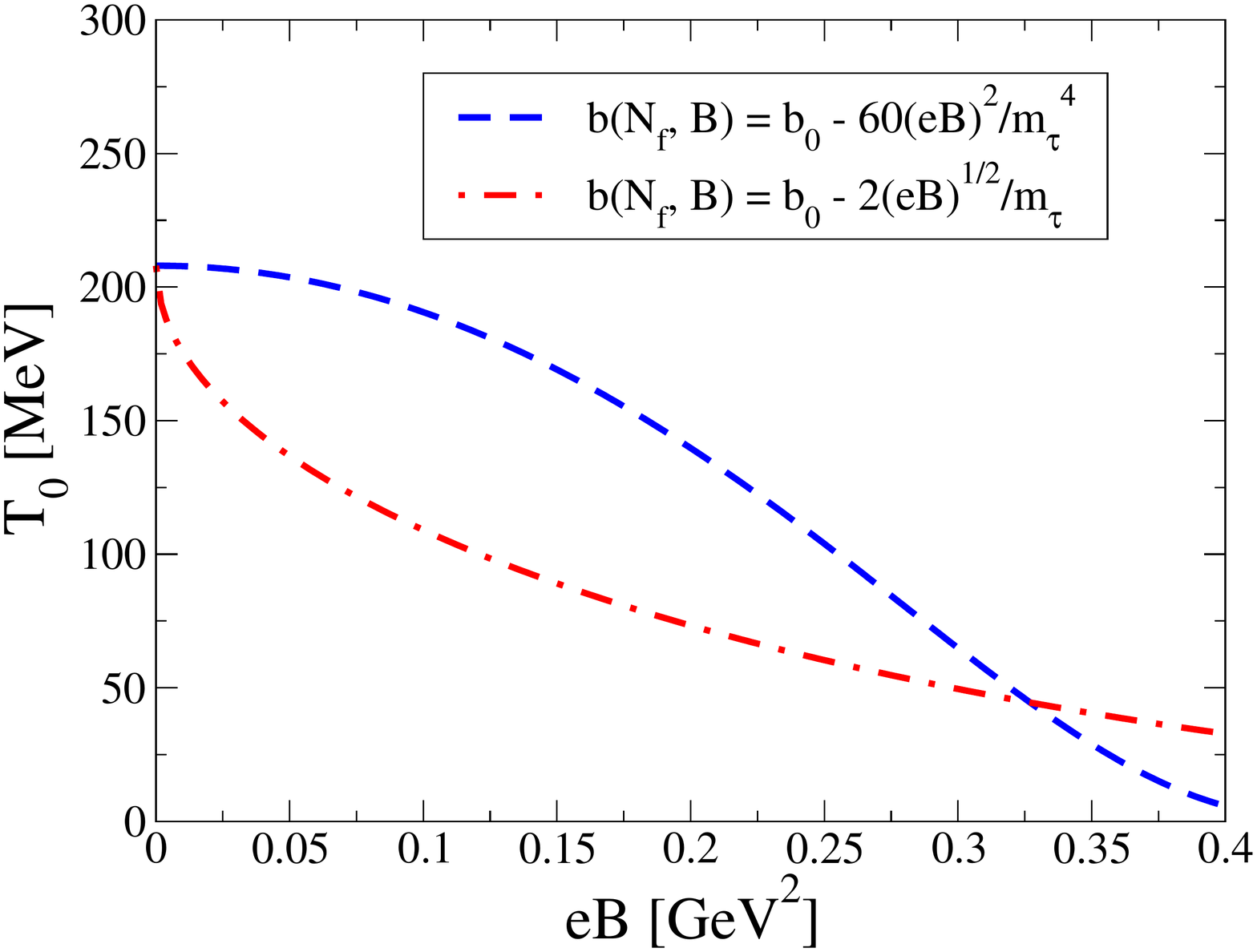}}
 \subfloat[]{\includegraphics[width=7cm]{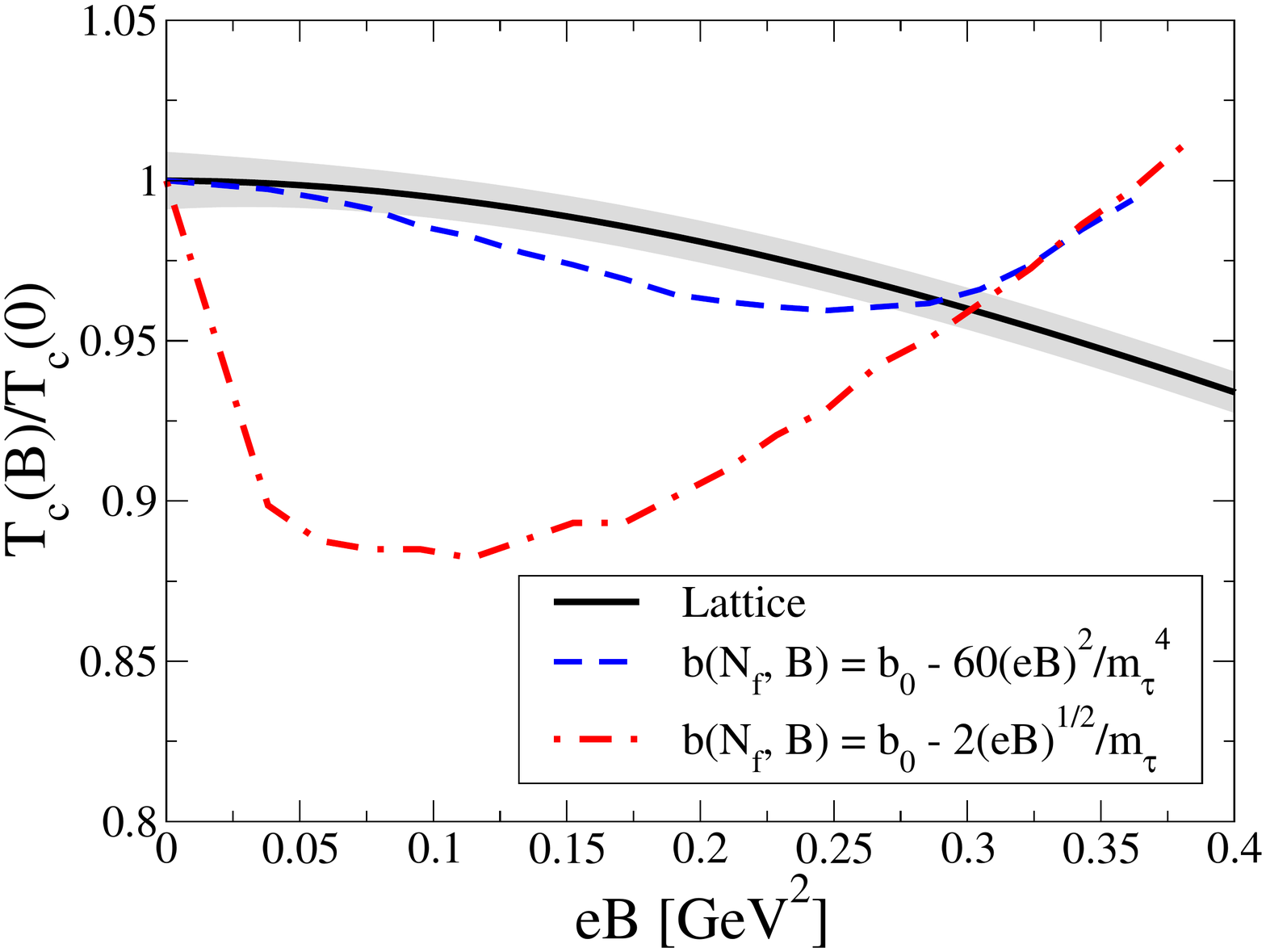}}
 \caption{(a) Value of the parameter $T_0$ for the two choices $b(N_f,0,B) = b_0 - 60(eB)^2/m_\tau^4$ 
 and $b(N_f,0,B) = b_0 - 2\sqrt{eB}/m_\tau$. 
 (b) The corresponding normalized phase diagram $T_c(B)/T_c(0)$, in comparison with lattice simulations 
 \cite{Bali:2011qj} (gray band).}
 \label{fig4:t0-of-B--and--tc-of-B}
\end{figure}

Let us provide explicit examples. The choices $b(N_f,0,B) = b_0 - 60(eB)^2/m_\tau^4$
and $b(N_f,0,B) = b_0 - 2\sqrt{eB}/m_\tau$, where $b_0:=(11N_c-2N_f)/6\pi=29/6\pi$, 
lead to the functions $T_0(B)$ shown in Fig. \ref{fig4:t0-of-B--and--tc-of-B}(a). 
The resulting normalized pseudo critical temperatures $T_c(B)/T_c(0)$ are 
shown\footnote{Our calculation ceases at $eB\sim0.3$GeV$^2$ due to numerical 
instabilities which arise when $T_0(B)$ is sufficiently small. Such instabilities are 
related to the Polyakov loop potential (\ref{eq:VPolyakov}), which is ill-defined at 
$T_0\rightarrow0$.} in Fig. \ref{fig4:t0-of-B--and--tc-of-B}(b) and compared to the 
(also normalized) lattice phase diagram of \cite{Bali:2011qj}. As advertised, the 
pseudo critical temperature goes down for low values of magnetic field, following 
the same trend as the lattice result from \cite{Bali:2011qj}, or maybe decreasing 
even faster. 
However, for moderate field intensities ($eB\lesssim0.3GeV^2\simeq15m_\pi^2$ 
or lower in the examples shown), the pseudo critical temperature rises
in spite of the very low values of $T_0(B)$ (see Fig. \ref{fig4:t0-of-B--and--tc-of-B}(a)). 
This behavior is due to the vacuum term Eq. (\ref{eq:VvacB}), which induces the 
magnetic catalysis in the vacuum and dominates over the thermal contribution 
Eq. (\ref{eq:VparaB}) for larger values of $B$.

\section{Conclusions}

We examined the behavior of the pseudo critical temperature for the
chiral transition in the QM and in the PQM model in the presence of a
constant magnetic field background by allowing for two natural
extensions of the parameter space: a $B$-dependent Yukawa coupling,
$g(B)$, in the quark-meson model and a $B$-dependent Polyakov loop
potential parameter, $T_0(B)$, in the PQM model.
 
We found that even strongly varying functions $g(B)$ and $T_0(B)$ are
not able to reproduce (not even qualitatively) the phase diagram
$T_c(B)$ from lattice simulations in the presence of a magnetic field
up to fields $eB\sim1$GeV$^2$. This robustness of the (P)QM model can
be attributed to the vacuum term, Eq. (\ref{eq:VvacB}), which is
responsible for magnetic catalysis in the vacuum and still dominates
the effective potential at finite temperature, even for moderate
values of the magnetic field intensity.

Even though our analysis has been performed at mean-field level, we
believe that our conclusions will not be changed if one considers
fluctuations (as studied, e.g., in Refs. \cite{Fukushima:2012xw} and 
\cite{Andersen:2013swa} for
fixed values of the model parameters). Likewise, there is no reason to
believe that other chiral models, such as the PNJL model, will behave
differently.

The physical mechanism for inverse magnetic catalysis recently 
proposed in \cite{Bruckmann:2013oba,Bruckmann:2013ufa} relies 
on a competition between valence and sea quarks. While the 
coupling of the magnetic field to valence quarks enhances 
magnetic catalysis, its interaction with sea quarks tends to 
decrease the chiral condensate, leading to inverse magnetic 
catalysis. In the context of chiral models, however, quarks 
are introduced as free quasiparticles, so that no degrees 
of freedom corresponding to sea quarks are present. We 
believe this can be an important, if not decisive, factor for 
the inability of chiral models to correctly describe the $T-B$ 
phase diagram. 

It seems clear that further comparisons between predictions from these
effective models and lattice data in systems containing other control
parameters besides the temperature are crucial to test the reliability
of those models. Besides the case with a magnetic field, these models
seem to face difficulties describing lattice data also when one
considers the dependence of the critical temperature on isospin and
quark masses \cite{Dumitru:2003cf,Fraga:2008be,Stiele:2013pma}.
Apparently, the presently used chiral models miss some important
physics to describe the properties of bulk QCD matter as extracted
from lattice simulations when introducing new control parameters, here
the magnetic field, even on a qualitative level. In view of this
circumstance, it seems that extrapolations of these chiral models to
regions of the QCD phase diagram not easily accessible by lattice
simulations have to be taken with caution before these basic
discrepancies are resolved.


\begin{acknowledgments}
  The authors thank Rainer Stiele for many valuable discussions during
  the initial stage of this work. The lattice data points of Fig. 
  \ref{fig4:t0-of-B--and--tc-of-B}(b) 
  were kindly provided by Gergely Endr\H{o}di. 
  BWM thanks the Institut f\"ur
  Theoretische Physik of the Goethe Universit\"at, Frankfurt, for
  their kind hospitality during a visit in which this project was
  initiated.  The work of ESF was financially supported by the
  Helmholtz International Center for FAIR within the framework of the
  LOEWE program (Landesoffensive zur Entwicklung
  Wissenschaftlich-\"Okonomischer Exzellenz) launched by the State of
  Hesse. The work of JSB was supported by the Hessian LOEWE initiative 
  through the Helmholtz International Center for FAIR (HIC for FAIR)
  and the Alliance Program of the Helmholtz Association (HA216/EMMI).
\end{acknowledgments}

\bibliographystyle{utphys}
\bibliography{references} 

\providecommand{\href}[2]{#2}\begingroup\raggedright\begin{thebibliography}{10}

\bibitem{Kharzeev:2013jha}
D.~Kharzeev, K.~Landsteiner, A.~Schmitt, and H.-U. Yee, ``{Strongly Interacting
  Matter in Magnetic Fields},''
\href{http://dx.doi.org/10.1007/978-3-642-37305-3}{{\em Lect.Notes Phys.}
  {\bfseries 871} (2013) 1--624}.

\bibitem{Fraga:2012rr}
E.~S. Fraga, ``{Thermal chiral and deconfining transitions in the presence of a
  magnetic background},''
  \href{http://dx.doi.org/10.1007/978-3-642-37305-3\_5}{{\em Lect.Notes Phys.}
  {\bfseries 871} (2013) 121--141},
\href{http://arxiv.org/abs/1208.0917}{{\ttfamily arXiv:1208.0917 [hep-ph]}}.

\bibitem{Gatto:2012sp}
R.~Gatto and M.~Ruggieri, ``{Quark Matter in a Strong Magnetic Background},''
  \href{http://dx.doi.org/10.1007/978-3-642-37305-3\_4}{{\em Lect.Notes Phys.}
  {\bfseries 871} (2013) 87--119},
\href{http://arxiv.org/abs/1207.3190}{{\ttfamily arXiv:1207.3190 [hep-ph]}}.

\bibitem{Agasian:2008tb}
N.~Agasian and S.~Fedorov, ``{Quark-hadron phase transition in a magnetic
  field},'' \href{http://dx.doi.org/10.1016/j.physletb.2008.04.050}{{\em
  Phys.Lett.} {\bfseries B663} (2008) 445--449},
\href{http://arxiv.org/abs/0803.3156}{{\ttfamily arXiv:0803.3156 [hep-ph]}}.

\bibitem{Fraga:2008qn}
E.~S. Fraga and A.~J. Mizher, ``{Chiral transition in a strong magnetic
  background},'' \href{http://dx.doi.org/10.1103/PhysRevD.78.025016}{{\em
  Phys.Rev.} {\bfseries D78} (2008) 025016},
\href{http://arxiv.org/abs/0804.1452}{{\ttfamily arXiv:0804.1452 [hep-ph]}}.

\bibitem{Menezes:2008qt}
D.~Menezes, M.~Benghi~Pinto, S.~Avancini, A.~Perez~Martinez, and
  C.~Providencia, ``{Quark matter under strong magnetic fields in the
  Nambu-Jona-Lasinio Model},''
  \href{http://dx.doi.org/10.1103/PhysRevC.79.035807}{{\em Phys.Rev.}
  {\bfseries C79} (2009) 035807},
\href{http://arxiv.org/abs/0811.3361}{{\ttfamily arXiv:0811.3361 [nucl-th]}}.

\bibitem{Boomsma:2009yk}
J.~K. Boomsma and D.~Boer, ``{The Influence of strong magnetic fields and
  instantons on the phase structure of the two-flavor NJL model},''
  \href{http://dx.doi.org/10.1103/PhysRevD.81.074005}{{\em Phys.Rev.}
  {\bfseries D81} (2010) 074005},
\href{http://arxiv.org/abs/0911.2164}{{\ttfamily arXiv:0911.2164 [hep-ph]}}.

\bibitem{Mizher:2010zb}
A.~J. Mizher, M.~Chernodub, and E.~S. Fraga, ``{Phase diagram of hot QCD in an
  external magnetic field: possible splitting of deconfinement and chiral
  transitions},'' \href{http://dx.doi.org/10.1103/PhysRevD.82.105016}{{\em
  Phys.Rev.} {\bfseries D82} (2010) 105016},
\href{http://arxiv.org/abs/1004.2712}{{\ttfamily arXiv:1004.2712 [hep-ph]}}.

\bibitem{Fukushima:2010fe}
K.~Fukushima, M.~Ruggieri, and R.~Gatto, ``{Chiral magnetic effect in the PNJL
  model},'' \href{http://dx.doi.org/10.1103/PhysRevD.81.114031}{{\em Phys.Rev.}
  {\bfseries D81} (2010) 114031},
\href{http://arxiv.org/abs/1003.0047}{{\ttfamily arXiv:1003.0047 [hep-ph]}}.

\bibitem{Gatto:2010pt}
R.~Gatto and M.~Ruggieri, ``{Deconfinement and Chiral Symmetry Restoration in a
  Strong Magnetic Background},''
  \href{http://dx.doi.org/10.1103/PhysRevD.83.034016}{{\em Phys.Rev.}
  {\bfseries D83} (2011) 034016},
\href{http://arxiv.org/abs/1012.1291}{{\ttfamily arXiv:1012.1291 [hep-ph]}}.

\bibitem{Kashiwa:2011js}
K.~Kashiwa, ``{Entanglement between chiral and deconfinement transitions under
  strong uniform magnetic background field},''
  \href{http://dx.doi.org/10.1103/PhysRevD.83.117901}{{\em Phys.Rev.}
  {\bfseries D83} (2011) 117901},
\href{http://arxiv.org/abs/1104.5167}{{\ttfamily arXiv:1104.5167 [hep-ph]}}.

\bibitem{Chatterjee:2011ry}
B.~Chatterjee, H.~Mishra, and A.~Mishra, ``{Vacuum structure and chiral
  symmetry breaking in strong magnetic fields for hot and dense quark
  matter},'' \href{http://dx.doi.org/10.1103/PhysRevD.84.014016}{{\em
  Phys.Rev.} {\bfseries D84} (2011) 014016},
\href{http://arxiv.org/abs/1101.0498}{{\ttfamily arXiv:1101.0498 [hep-ph]}}.

\bibitem{Andersen:2011ip}
J.~O. Andersen and R.~Khan, ``{Chiral transition in a magnetic field and at
  finite baryon density},''
  \href{http://dx.doi.org/10.1103/PhysRevD.85.065026}{{\em Phys.Rev.}
  {\bfseries D85} (2012) 065026},
\href{http://arxiv.org/abs/1105.1290}{{\ttfamily arXiv:1105.1290 [hep-ph]}}.

\bibitem{Skokov:2011ib}
V.~Skokov, ``{Phase diagram in an external magnetic field beyond a mean-field
  approximation},'' \href{http://dx.doi.org/10.1103/PhysRevD.85.034026}{{\em
  Phys.Rev.} {\bfseries D85} (2012) 034026},
\href{http://arxiv.org/abs/1112.5137}{{\ttfamily arXiv:1112.5137 [hep-ph]}}.

\bibitem{Andersen:2012bq}
J.~O. Andersen and A.~Tranberg, ``{The Chiral transition in a magnetic
  background: Finite density effects and the functional renormalization
  group},'' \href{http://dx.doi.org/10.1007/JHEP08(2012)002}{{\em JHEP}
  {\bfseries 1208} (2012) 002},
\href{http://arxiv.org/abs/1204.3360}{{\ttfamily arXiv:1204.3360 [hep-ph]}}.

\bibitem{Fukushima:2012xw}
K.~Fukushima and J.~M. Pawlowski, ``{Magnetic catalysis in hot and dense quark
  matter and quantum fluctuations},''
  \href{http://dx.doi.org/10.1103/PhysRevD.86.076013}{{\em Phys.Rev.}
  {\bfseries D86} (2012) 076013},
\href{http://arxiv.org/abs/1203.4330}{{\ttfamily arXiv:1203.4330 [hep-ph]}}.

\bibitem{Fraga:2012fs}
E.~S. Fraga and L.~F. Palhares, ``{Deconfinement in the presence of a strong
  magnetic background: an exercise within the MIT bag model},''
  \href{http://dx.doi.org/10.1103/PhysRevD.86.016008}{{\em Phys.Rev.}
  {\bfseries D86} (2012) 016008},
\href{http://arxiv.org/abs/1201.5881}{{\ttfamily arXiv:1201.5881 [hep-ph]}}.

\bibitem{Strickland:2012vu}
M.~Strickland, V.~Dexheimer, and D.~Menezes, ``{Bulk Properties of a Fermi Gas
  in a Magnetic Field},''
  \href{http://dx.doi.org/10.1103/PhysRevD.86.125032}{{\em Phys.Rev.}
  {\bfseries D86} (2012) 125032},
\href{http://arxiv.org/abs/1209.3276}{{\ttfamily arXiv:1209.3276 [nucl-th]}}.

\bibitem{Fukushima:2012kc}
K.~Fukushima and Y.~Hidaka, ``{Magnetic Catalysis vs Magnetic Inhibition},''
  \href{http://dx.doi.org/10.1103/PhysRevLett.110.031601}{{\em Phys.Rev.Lett.}
  {\bfseries 110} (2013) 031601},
\href{http://arxiv.org/abs/1209.1319}{{\ttfamily arXiv:1209.1319 [hep-ph]}}.

\bibitem{Ferrari:2012yw}
G.~N. Ferrari, A.~F. Garcia, and M.~B. Pinto, ``{Chiral Transition Within
  Effective Quark Models Under Magnetic Fields},''
  \href{http://dx.doi.org/10.1103/PhysRevD.86.096005}{{\em Phys.Rev.}
  {\bfseries D86} (2012) 096005},
\href{http://arxiv.org/abs/1207.3714}{{\ttfamily arXiv:1207.3714 [hep-ph]}}.

\bibitem{Allen:2013eha}
P.~G. Allen and N.~N. Scoccola, ``{Phase diagram of strongly interacting matter
  under strong magnetic fields},''
\href{http://arxiv.org/abs/1307.4070}{{\ttfamily arXiv:1307.4070 [hep-ph]}}.

\bibitem{Fu:2013ica}
W.-j. Fu, ``{Fluctuations and correlations of hot QCD matter in an external
  magnetic field},''
\href{http://arxiv.org/abs/1306.5804}{{\ttfamily arXiv:1306.5804 [hep-ph]}}.

\bibitem{Andersen:2012jf}
J.~O. Andersen and A.~A. Cruz, ``{Two-color QCD in a strong magnetic field: The
  role of the Polyakov loop},''
  \href{http://dx.doi.org/10.1103/PhysRevD.88.025016}{{\em Phys.Rev.}
  {\bfseries D88} (2013) 025016},
\href{http://arxiv.org/abs/1211.7293}{{\ttfamily arXiv:1211.7293 [hep-ph]}}.

\bibitem{Andersen:2012zc}
J.~O. Andersen, ``{Chiral perturbation theory in a magnetic background -
  finite-temperature effects},''
  \href{http://dx.doi.org/10.1007/JHEP10(2012)005}{{\em JHEP} {\bfseries 1210}
  (2012) 005},
\href{http://arxiv.org/abs/1205.6978}{{\ttfamily arXiv:1205.6978 [hep-ph]}}.

\bibitem{Avancini:2012ee}
S.~S. Avancini, D.~P. Menezes, M.~B. Pinto, and C.~Providencia, ``{The QCD
  Critical End Point Under Strong Magnetic Fields},''
  \href{http://dx.doi.org/10.1103/PhysRevD.85.091901}{{\em Phys.Rev.}
  {\bfseries D85} (2012) 091901},
\href{http://arxiv.org/abs/1202.5641}{{\ttfamily arXiv:1202.5641 [hep-ph]}}.

\bibitem{Andersen:2012dz}
J.~O. Andersen, ``{Thermal pions in a magnetic background},''
  \href{http://dx.doi.org/10.1103/PhysRevD.86.025020}{{\em Phys.Rev.}
  {\bfseries D86} (2012) 025020},
\href{http://arxiv.org/abs/1202.2051}{{\ttfamily arXiv:1202.2051 [hep-ph]}}.

\bibitem{Costa:2013zca}
P.~Costa, M.~Ferreira, H.~Hansen, D.~P. Menezes, and C.~Providência, ``{Phase
  transition and CEP driven by an external magnetic field in asymmetric quark
  matter},''
\href{http://arxiv.org/abs/1307.7894}{{\ttfamily arXiv:1307.7894 [hep-ph]}}.

\bibitem{Allen:2013lda}
P.~G. Allen and N.~N. Scoccola, ``{Quark matter under strong magnetic fields in
  SU(2) NJL-type models: parameter dependence of the cold dense matter phase
  diagram},''
\href{http://arxiv.org/abs/1309.2258}{{\ttfamily arXiv:1309.2258 [hep-ph]}}.

\bibitem{Ballon-Bayona:2013cta}
A.~Ballon-Bayona, ``{Holographic deconfinement transition in the presence of a
  magnetic field},''
\href{http://arxiv.org/abs/1307.6498}{{\ttfamily arXiv:1307.6498 [hep-th]}}.

\bibitem{Callebaut:2013ria}
N.~Callebaut and D.~Dudal, ``{On the transition temperature(s) of magnetized
  two-flavour holographic QCD},''
  \href{http://dx.doi.org/10.1103/PhysRevD.87.106002}{{\em Phys.Rev.}
  {\bfseries D87} (2013) 106002},
\href{http://arxiv.org/abs/1303.5674}{{\ttfamily arXiv:1303.5674 [hep-th]}}.

\bibitem{Ferreira:2013tba}
M.~Ferreira, P.~Costa, D.~P. Menezes, C.~Providência, and N.~Scoccola,
  ``{Deconfinement and chiral restoration within the SU(3) PNJL and EPNJL
  models in an external magnetic field},''
\href{http://arxiv.org/abs/1305.4751}{{\ttfamily arXiv:1305.4751 [hep-ph]}}.

\bibitem{Ruggieri:2013cya}
M.~Ruggieri, M.~Tachibana, and V.~Greco, ``{Renormalized vs Nonrenormalized
  Chiral Transition in a Magnetic Background},''
\href{http://arxiv.org/abs/1305.0137}{{\ttfamily arXiv:1305.0137 [hep-ph]}}.

\bibitem{Chernodub:2010qx}
M.~Chernodub, ``{Superconductivity of QCD vacuum in strong magnetic field},''
  \href{http://dx.doi.org/10.1103/PhysRevD.82.085011}{{\em Phys.Rev.}
  {\bfseries D82} (2010) 085011},
\href{http://arxiv.org/abs/1008.1055}{{\ttfamily arXiv:1008.1055 [hep-ph]}}.

\bibitem{Chernodub:2011mc}
M.~Chernodub, ``{Spontaneous electromagnetic superconductivity of vacuum in
  strong magnetic field: evidence from the Nambu--Jona-Lasinio model},''
  \href{http://dx.doi.org/10.1103/PhysRevLett.106.142003}{{\em Phys.Rev.Lett.}
  {\bfseries 106} (2011) 142003},
\href{http://arxiv.org/abs/1101.0117}{{\ttfamily arXiv:1101.0117 [hep-ph]}}.

\bibitem{D'Elia:2010nq}
M.~D'Elia, S.~Mukherjee, and F.~Sanfilippo, ``{QCD Phase Transition in a Strong
  Magnetic Background},''
  \href{http://dx.doi.org/10.1103/PhysRevD.82.051501}{{\em Phys.Rev.}
  {\bfseries D82} (2010) 051501},
\href{http://arxiv.org/abs/1005.5365}{{\ttfamily arXiv:1005.5365 [hep-lat]}}.

\bibitem{D'Elia:2011zu}
M.~D'Elia and F.~Negro, ``{Chiral Properties of Strong Interactions in a
  Magnetic Background},''
  \href{http://dx.doi.org/10.1103/PhysRevD.83.114028}{{\em Phys.Rev.}
  {\bfseries D83} (2011) 114028},
\href{http://arxiv.org/abs/1103.2080}{{\ttfamily arXiv:1103.2080 [hep-lat]}}.

\bibitem{Bali:2011qj}
G.~Bali, F.~Bruckmann, G.~Endrodi, Z.~Fodor, S.~Katz, {\em et~al.}, ``{The QCD
  phase diagram for external magnetic fields},''
  \href{http://dx.doi.org/10.1007/JHEP02(2012)044}{{\em JHEP} {\bfseries 1202}
  (2012) 044},
\href{http://arxiv.org/abs/1111.4956}{{\ttfamily arXiv:1111.4956 [hep-lat]}}.

\bibitem{Bali:2012zg}
G.~Bali, F.~Bruckmann, G.~Endrodi, Z.~Fodor, S.~Katz, {\em et~al.}, ``{QCD
  quark condensate in external magnetic fields},''
  \href{http://dx.doi.org/10.1103/PhysRevD.86.071502}{{\em Phys.Rev.}
  {\bfseries D86} (2012) 071502},
\href{http://arxiv.org/abs/1206.4205}{{\ttfamily arXiv:1206.4205 [hep-lat]}}.

\bibitem{Bruckmann:2013oba}
F.~Bruckmann, G.~Endrodi, and T.~G. Kovacs, ``{Inverse magnetic catalysis and
  the Polyakov loop},'' \href{http://dx.doi.org/10.1007/JHEP04(2013)112}{{\em
  JHEP} {\bfseries 1304} (2013) 112},
\href{http://arxiv.org/abs/1303.3972}{{\ttfamily arXiv:1303.3972 [hep-lat]}}.

\bibitem{Fraga:2012ev}
E.~S. Fraga, J.~Noronha, and L.~F. Palhares, ``{Large Nc Deconfinement
  Transition in the Presence of a Magnetic Field},''
  \href{http://dx.doi.org/10.1103/PhysRevD.87.114014}{{\em Phys.Rev.}
  {\bfseries D87} (2013) 114014},
\href{http://arxiv.org/abs/1207.7094}{{\ttfamily arXiv:1207.7094 [hep-ph]}}.

\bibitem{Blaizot:2012sd}
J.-P. Blaizot, E.~S. Fraga, and L.~F. Palhares, ``{Effect of quark masses on
  the QCD presssure in a strong magnetic background},''
  \href{http://dx.doi.org/10.1016/j.physletb.2013.04.004}{{\em Phys.Lett.}
  {\bfseries B722} (2013) 167--171},
\href{http://arxiv.org/abs/1211.6412}{{\ttfamily arXiv:1211.6412 [hep-ph]}}.

\bibitem{Roessner_08}
S.~{R{\"o}{\ss}ner}, T.~{Hell}, C.~{Ratti}, and W.~{Weise}, ``{The chiral and
  deconfinement crossover transitions: PNJL model beyond mean field},''
  \href{http://dx.doi.org/10.1016/j.nuclphysa.2008.10.006}{{\em Nucl. Phys. A}
  {\bfseries 814} (2008) 118--143},
  \href{http://arxiv.org/abs/0712.3152}{{\ttfamily arXiv:0712.3152 [hep-ph]}}.

\bibitem{PDG_12}
{J. Beringer and Particle Data Group}, ``{Review of Particle Physics},''
  \href{http://dx.doi.org/10.1103/PhysRevD.86.010001}{{\em Phys. Rev. D}
  {\bfseries 86} no.~1, (2012) 010001}.

\bibitem{Schaefer_07}
B.~{Schaefer}, J.~M. {Pawlowski}, and J.~{Wambach}, ``{Phase structure of the
  Polyakov-quark-meson model},''
  \href{http://dx.doi.org/10.1103/PhysRevD.76.074023}{{\em Phys. Rev. D}
  {\bfseries 76} no.~7, (2007) 074023--+},
  \href{http://arxiv.org/abs/0704.3234}{{\ttfamily arXiv:0704.3234 [hep-ph]}}.

\bibitem{Karsch:2000kv}
F.~Karsch, E.~Laermann, and A.~Peikert, ``{Quark mass and flavor dependence of
  the QCD phase transition},''
  \href{http://dx.doi.org/10.1016/S0550-3213(01)00200-0}{{\em Nucl.Phys.}
  {\bfseries B605} (2001) 579--599},
\href{http://arxiv.org/abs/hep-lat/0012023}{{\ttfamily arXiv:hep-lat/0012023
  [hep-lat]}}.

\bibitem{Bruckmann:2013ufa}
F.~Bruckmann, G.~Endrodi, and T.~G. Kovacs, ``{Inverse magnetic catalysis in
  QCD},''
\href{http://arxiv.org/abs/1311.3178}{{\ttfamily arXiv:1311.3178 [hep-lat]}}.

\bibitem{Andersen:2013swa}
J.~O. Andersen, W.~R. Naylor, and A.~Tranberg, ``{Chiral and deconfinement
  transitions in a magnetic background using the functional renormalization
  group with the Polyakov loop},''
\href{http://arxiv.org/abs/1311.2093}{{\ttfamily arXiv:1311.2093 [hep-ph]}}.

\bibitem{Dumitru:2003cf}
A.~Dumitru, D.~Roder, and J.~Ruppert, ``{The Quark mass dependence of T(c) in
  QCD: Working up from m = 0 or down from m = infinity?},''
  \href{http://dx.doi.org/10.1103/PhysRevD.70.074001}{{\em Phys.Rev.}
  {\bfseries D70} (2004) 074001},
\href{http://arxiv.org/abs/hep-ph/0311119}{{\ttfamily arXiv:hep-ph/0311119
  [hep-ph]}}.

\bibitem{Fraga:2008be}
E.~Fraga, L.~Palhares, and C.~Villavicencio, ``{Quark mass and isospin
  dependence of the deconfining critical temperature},''
  \href{http://dx.doi.org/10.1103/PhysRevD.79.014021}{{\em Phys.Rev.}
  {\bfseries D79} (2009) 014021},
\href{http://arxiv.org/abs/0810.1060}{{\ttfamily arXiv:0810.1060 [hep-ph]}}.

\bibitem{Stiele:2013pma}
R.~Stiele, E.~S. Fraga, and J.~Schaffner-Bielich, ``{Thermodynamics of
  (2+1)-flavor strongly interacting matter at nonzero isospin},''
\href{http://arxiv.org/abs/1307.2851}{{\ttfamily arXiv:1307.2851 [hep-ph]}}.

\end{thebibliography}\endgroup

\end{document}